\documentclass[10pt]{article}
\usepackage{graphics}
\makeatletter
\def\fakecaptype#1{\def\@captype{#1}}
\makeatother

\makeatletter 
\def\eqnarray{%
   \stepcounter{equation}%
   \def\@currentlabel{\p@equation\theequation}%
   \global\@eqnswtrue
   \m@th
   \global\@eqcnt\z@
   \tabskip\@centering
   \let\\\@eqncr
   $$\everycr{}\halign to\displaywidth\bgroup
       \hskip\@centering$\displaystyle\tabskip\z@skip{##}$\@eqnsel
      &\global\@eqcnt\@ne \hfil${{}##{}}$\hfil
      &\global\@eqcnt\tw@
         $\displaystyle{##}$\hfil\tabskip\@centering
      &\global\@eqcnt\thr@@ \hb@xt@\z@\bgroup\hss##\egroup
         \tabskip\z@skip
      \cr
}
\newenvironment{subequations}{%
  \refstepcounter{equation}%
  \mathchardef\c@mainequation\c@equation
  \protected@edef\themainequation{\theequation}%
  \let\theequation\thesubequation
  \global\c@equation\z@
  }{%
  \global\c@equation\c@mainequation
  \global\@ignoretrue
  }
\newcommand{\thesubequation}{\themainequation\alph{equation}}

\newcommand*{\@lab@subeqnarray}[2]{#1{#2}\eqnarray}
\@addtoreset{equation}{chapter}
\def\theequation{\@arabic\c@section.\@arabic\c@equation}
\makeatother
\newcommand\eref[1]{(\ref{#1})}
\newcommand\n[3]{\ensuremath{{#1}\geq{#2}\geq{#3}}}
\def\d{{\rm d}}


\begin{document}

\begin{center}

{\bf \Large A charged rotating cylindrical shell\footnote[1]
{Suggested running head: 
Charged rotating cylindrical shell}}\\[3mm]

J. Ger\v{s}l\footnote[2]
{Department of General Physics, Faculty of Science,\\ Masaryk 
University, Kotl\'{a}\v{r}sk\'{a} 2, 611 37  Brno, 
Czech Republic}
\footnote[3]
{E-mail: {\tt janger@physics.muni.cz}}, 
P. Klep\'a\v{c}\footnote[4]
{Institute of Theoretical Physics and 
Astrophysics, Faculty of Science,\\ Masaryk 
University, Kotl\'{a}\v{r}sk\'{a} 2, 611 37  Brno, 
Czech Republic}
\footnote[5]
{E-mail: {\tt klepac@physics.muni.cz}} 
and J. Horsk\'y$^{\ 4}$
\footnote[6]{E-mail: 
{\tt horsky@physics.muni.cz}}\\[2mm]
\end{center} 

\noindent
{\small
We give an example of a spacetime having an infinite thin rotating
cylindrical shell constituted by a charged perfect fluid as a source.
As the interior of the shell the Bonnor--Melvin universe is
considered, while its exterior is represented by Datta--Raychaudhuri
spacetime. We discuss the energy conditions and we show that our spacetime
contains closed timelike curves. Trajectories of charged test
particles both inside and outside the cylinder are also examined.
Expression for the angular velocity of a circular motion inside
the cylinder is given.\\

}

\newpage

\section{Introduction}

Cylindrically symmetric sources of gravitational field have been
intensively studied during the whole development of general
relativity. Although they are unbounded and can not represent real
objects creating gravitational field, they constitute framework for
investigation of spacetimes with a high degree of symmetry in the
field of exact as well as numerical solutions to the Einstein equations
\cite{{kramer},{bonnor},{MacCallum}}. 
Cylindrically symmetric sources have become
of a great significance in context of relativistic cosmology
\cite{{bonnor},{Pavel}}  and they provide an important tool for
examining dynamical models, e.g. in cases with presence of
gravitational waves \cite{wave}.  They have been connected with
considerations taking quantum gravity and probe trajectories in
context of string theory into account as well \cite{Wright}.

A number of papers have been concerned with a rotating cylinder in
general relativity (\cite{{bonnor},{bonnor2}} and references therein).
In particular, both rotating and nonrotating cylindrical thin shells have
been studied \cite{{Embacher},{wang}}. 
In the works \cite{Zofka} various
kinds of shell sources for static Levi--Civita and Lewis spacetime
respectively have been discussed. Charged generalizations of the Levi--Civita 
spacetime and their shell sources have been studied in \cite{Santos}.

The outline of the paper is the following. Section \ref{sec2}
characterizes the physical set up -- matching the Bonnor--Melvin
magnetic universe \cite{bonnor3} to solution of Datta and Raychaudhuri
\cite{datta}. In section \ref{sec3}, by using of the Israel--Kucha\v{r}
junction conditions \cite{{israel},{Kuchar}}, we put forward a spacetime with an
infinite rotating cylindrical shell, built up from a charged perfect
fluid acting as a source of the gravitational and electromagnetic field.  Afterward, in
section \ref{properties}, we discuss some attributes of the solution,
in particular a question of the chronology violation in the spacetime
and the energy conditions. Finally in section \ref{motion} the
question is addressed what is a qualitative behavior of trajectories
of test particles, either charged or uncharged. 

\section{\label{sec2}An interior and an exterior spacetime and a shell}

The goal of the paper is to obtain a spacetime having a cylindrical
shell as its source. The spacetime arises by matching a conveniently
chosen interior spacetime to a suitably chosen exterior one. From
reasons given bellow the interior portion is to be constituted by (a
portion of) the Bonnor--Melvin magnetic universe
\cite{{bonnor3},{kramer}} while as the exterior one we have chosen
solution of Datta and Raychaudhuri \cite{datta}. It is worth to note
that these two spacetimes share the same symmetries, namely they are
cylindrically symmetric and stationary.

The Bonnor--Melvin magnetic universe (BM) describes a static 
electrovacuum spacetime and we restrict ourselves to values of radial 
coordinate $r\in(0,r_B]$. The metric and the electromagnetic 
potential of this spacetime have the form\footnote{In the paper the
natural units has been used, i.e. $c=G=1$ and $\mu_0=4\pi$.} 
\begin{eqnarray}
\d s^2\!=\!-(1\!-\!K^2r^2)^{-2}\d t^2\!+(1\!-\!K^2r^2)r^2\d \phi ^2
+(1\!-\!K^2r^2)^{-2}\d z^2\!+\!(1\!-\!K^2r^2)^{-5}\d r^2 ,\nonumber\\
A=Kr^2\d \phi \hspace{5cm}\label{MC}
\end{eqnarray}
with $K$ being a constant, $t\in (-\infty,\infty),\ z\in (-\infty,\infty)$ and $\phi$ being $2\pi$-periodic angular coordinate. 
Correctness of the signature requires $r_B<\frac{1}{| K|}$. 
The choice of BM spacetime is motivated by the following facts. If we
eliminate the gravitational coupling of an electromagnetic field 
in this spacetime by setting the gravitational constant equal to zero,
we get the Minkowski spacetime with a homogeneous magnetic field of
the magnitude $2K$ pointing in $z$ direction. But this is exactly the
interior of a rotating charged cylindrical shell in special relativity theory. 
Furthermore, BM spacetime has no singularity on the rotation axis. In
addition, it satisfies the elementary flatness condition.  
  
The exterior of the cylinder is constituted by Datta and Raychaudhuri
(DR) solution for values of radial coordinate $r\in[r_D,\infty)$. 
The metric and the electromagnetic potential of DR spacetime have the form
\begin{eqnarray}
\d s^2&=&(-U^2(4r^2+\lambda r\ln(r))+2U\Omega r )\d t^2\!+
(-H^2(4r^2\!+\lambda r\ln(r))\!+\!2HDr)\d \phi ^2+\nonumber\\
&+&2(-UH(4r^2+\lambda r\ln(r))+(\Omega H+UD)r)\d t\d \phi \nonumber\\
&+&V^2r^{-\frac{1}{2}}\d z^2+r^{-\frac{1}{2}}\d r^2\ ,\label{MCmet}\\
A&=&-r (U\d t+H\d \phi)\ ,\label{MCpot}
\end{eqnarray}
where $U,H,\Omega ,D,V,\lambda$ are constants. The coordinates 
$t, z$ have the same range as in BM and $\phi$ is assumed 
to be $2\pi$--periodic angular coordinate again. 

DR is a non--static spacetime with null electromagnetic field 
($F_{\mu\nu}F^{\mu\nu}=0$), which contains 
a radial electric field and a magnetic field pointing in $z$ direction.
DR solution is a charged generalization of Van Stockum
solution (see bellow). The metric and the
electromagnetic potential \eref{MCmet} and \eref{MCpot}
can be locally obtained from the following metric and electromagnetic 
potential 
\begin{eqnarray}\label{MC_puv}
\d s^2&=&-(4r^2+\lambda r\ln r)\d \bar{t}^2+2r\d \bar{t}\d \bar{\phi}
+r^{-\frac{1}{2}}\d \bar{z}^2+r^{-\frac{1}{2}}\d r^2\ ,\\
A&=&-r \d \bar{t}
\end{eqnarray}
by the linear transformation  
\begin{equation}\label{tra}\left(\begin{array}{c}t\\\phi\\z\end{array}\right)=
\left(\begin{array}{ccc}U&H&0\\
\Omega&D&0\\
0&0&V\end{array}\right)^{-1}
\left(\begin{array}{c}\bar{t}\\\bar{\phi}\\\bar{z}\end{array}\right).\end{equation}
The transformation matrix \eref{tra} have to be regular. If the coordinates with 
bars are used in spacetime 
\eref{MCmet}, \eref{MCpot}, then $\bar{\phi}$ isn't angular generally.
    
The boundary between BM and DR spacetimes is given by the equation
$r=r_B$ in BM and by the equation $r=r_D$ in DR. These 
hypersurfaces are joined by identification of the points 
with the same coordinates $t,\phi ,z$. Let us call $T,\ \Phi,\ Z$ the
coordinates that arise from the coordinate functions $t,\ \phi,\ z$ by 
restriction their domain to the boundary. 

Let a charged rotating cylindrical shell built up from a perfect fluid
be situated at the boundary. The shell surface energy -- momentum tensor 
has the form (\cite{Kuchar})
\footnote{Possible values of indexes $a,b$ are $T,\ \Phi,\ Z$.}
\begin{equation}t_{ab}=(p+\rho)u_{a}u_{b}+pg_{ab},\end{equation}
with $\rho$ being rest surface mass density, $p$ rest surface pressure 
on the shell and $g_{ab}$ being the components 
of the induced metric at the boundary. The values of $\rho,\ p$ 
do not depend on a position at the shell. 
Particles of the shell move with the 4--velocity 
\begin{equation}\label{4vel}
\mathbf{u}=\frac{\frac{\partial _T}{\sqrt{-g_{TT}}}+v\frac{\partial
_{\Phi}}{\sqrt{g_{\Phi\Phi}}}} {\sqrt{1-v^2}}\ ,\end{equation}
where $v$ is the velocity of the shell particles in 
$\phi$ direction as measured by observers moving on curves 
$\{\Phi,\ Z\}={\rm const}$. The condition $| v| <1$ 
must be satisfied in order that $\mathbf{u}$ be timelike.

\section{\label{sec3} Junction conditions and their solution}

We denote $g^+$ and $g^-$ the induced metric at the boundary 
from side of DR spacetime and BM spacetime respectively.
Similarly, let us denote $k^+$ and $k^-$ the external curvature of the
boundary with respect to outward normal vector field. Israel 
junction conditions for gravitational field \cite{israel} then
become\footnote{In the paper the sign conventions of \cite{MTW} have
been used.}
\begin{equation}\label{Is}g_{ab}^+=g_{ab}^-\ ,\qquad 
k^+_{ab}-k^-_{ab}=8\pi (t_{ab}-\frac{1}{2}tg_{ab}),\end{equation}
where $g_{ab}:=g_{ab}^+=g_{ab}^-$.
Furthermore let $\sigma$ be the rest surface charge density which does
not depend on a position at the shell, 
$\mathbf{s}=\sigma\mathbf{u}$ be the surface current
density of the shell, $\mathbf{n}$ outward normal to the boundary and 
$F^+$ and $F^-$ the electromagnetic field tensor in DR and BM spacetime. 
Kucha\v{r} junction conditions for the electromagnetic field \cite{Kuchar} have the form 
\begin{equation}\label{Ku}F^+(\partial _a,\mathbf{n})-F^-(\partial _a,\mathbf{n})=4\pi s_{a},\;\;\;\; F^+(\partial _a ,\partial _b)=
F^-(\partial _a ,\partial _b).\end{equation}
The conditions \eref{Is}, \eref{Ku} lead to a system of ten algebraic equations for thirteen unknowns in our case. 
By solving them one can express quantities 
$r_D,U,\Omega ,H,D, V,\lambda ,p,\rho ,\sigma$ in terms of 
functions of $K,r_B,v,h$, where $h=\pm 1$ arises from solving of a
certain quadratic equation. These functions take the form 
\begin{subequations}
\begin{eqnarray}
r_D^{-\frac{3}{4}}&=&\frac{4}{3}\frac{1-4K^2r_B^2}{r_BX^{-\frac{3}{2}}(1-hv)}\
,\label{r_D}\\
hU&=&\frac{2 r_D^{-\frac{1}{4}}KX}{1-hv}\ ,\label{hU}\\
H&=&-\frac{2r_B r_D^{-\frac{1}{4}}KX^{\frac{5}{2}}}{1-hv}\ ,\label{H}\\
h\Omega &=&-\frac{1-4K^2r_B^2}{3Kr_B
X^{\frac{3}{2}}}+\frac{hv (1-4K^2r_B^2)}{2Kr_BX^{\frac{3}{2}}(1-hv)}
\ln r_D\nonumber\\ 
&-&\frac{3Kr_BX^{-\frac{1}{2}}}{1-4K^2r_B^2}(\ln r_D -1)\ ,\label{h_Omega}\\
D&=&-\frac{2}{3K}(1-4K^2r_B^2)-r_BX^{\frac{3}{2}}h\Omega \ ,\label{D}\\
V&=&r_D^{\frac{1}{4}}X^{-1}\ ,\label{V}\\
\lambda &=&\frac{1}{hU}\left
(\frac{hv (1-4K^2r_B^2)}{Kr_BX^{\frac{3}{2}}(1-hv)}
-\frac{6Kr_BX^{-\frac{1}{2}}}{1-4K^2r_B^2}\right)\ ,\label{lambda}\\
16\pi\rho &=&r_D^{-\frac{3}{4}}+8K^2r_BX^{\frac{3}{2}}\ ,\label{rho}\\
16\pi(p+\rho)&=&\frac{2X^{\frac{3}{2}}(1-4K^2r_B^2)(1+hv)}{r_B(1-hv)}\
,\label{p}\\
4\pi h\sigma &=&-\frac{2KX^2}{1-hv}(1-v^2)^{\frac{1}{2}}\ ,\label{sigma}
\end{eqnarray}
\end{subequations}
where $X:=1-K^2r_B^2$. The condition for correct signature of \eref{MC} guarantees
that $X>0$. Moreover, the inequalities
\begin{equation}\mid Kr_B\mid <\frac{1}{2}\ ,\qquad K\neq 0\
\label{Kr_B}\end{equation}
have to be satisfied, to get positive right side of \eref{r_D} and to prevent infinities. The regularity of \eref{tra} requires 
$(UD-H\Omega)V=\frac{4}{3}\ \frac{1-4K^2r_B^2}{v-h}\neq 0$. This is satisfied, because \eref{Kr_B} holds. Note that the 
line element \eref{MCmet} remains unaffected by a transformation $v\rightarrow -v,\ \phi\rightarrow -\phi ,\  h\rightarrow 
-h$ instead of expected $v\rightarrow -v,\ \phi\rightarrow -\phi $. But we can redefine $\bar{h}=h\ \rm{sign}$$(v)$ 
for $v\neq 0$. Then $hv=\bar{h}|v|$ holds and keeping $\bar{h}$ unchanged 
we get the expected symmetry. A connection 
between $\bar{h}$ and physical properties of the obtained spacetime can be seen for example from a relation
$\bar{h}=-{\rm sign}(s_{\Phi}/B_Z)$ (for $s_\Phi\neq 0$), where $s_{\Phi}$ is $\Phi$--component of the
surface current density and $B_Z$ is $Z$--component of the magnetic
field at the shell taken from an arbitrary side of the shell. Thus a change 
of $\bar{h}$ implies a change of a direction in which the
magnetic field points out near the shell with respect to the current
on the shell.  

The uncharged spacetime can be obtained from the charged one by taking a limit $K\rightarrow 0$. Inside the shell it 
leads to the Minkowski spacetime. The same limit performed outside the shell yields 
\[\d s^2=-{\rm sign}(hv)r\ln(r)\d {\tilde t}^2+2r\d{\tilde \phi}\d{\tilde
t}+r^{-\frac12}\left(\d {\tilde z}^2+\d r^2\right)\ ,\]
which coincides with the Van Stockum metric \cite{kramer}.
 
\section{\label{properties}Physical properties of the matched spacetime}

\subsection{Energy conditions\label{energy_conditions}}

As one can see from \eref{rho} and \eref{p}, 
the weak energy condition, $\rho\geq 0, \rho +p\geq 0$, is satisfied for all allowed values of the free parameters. 

The strong energy condition, $p+\rho\geq 0$, $\rho +2p\geq 0$, gives the following restriction of the free parameters 
\begin{equation}\label{strong}
K^2r_B^2\leq\frac{1}{2}\ \frac{2+3hv}{7+3hv}\ .\end{equation}
The dominant energy condition is equivalent to the weak energy condition 
plus the inequality $\rho -p\geq 0$ which reads
\begin{equation}\label{dominant}
K^2r_B^2\geq\frac{1}{4}\ \frac{3hv-1}{5-3hv}\ .\end{equation}

\subsection{\label{charge_density}Linear charge density} 

\begin{figure}[t]
\begin{center}
\scalebox{0.6}{{\includegraphics{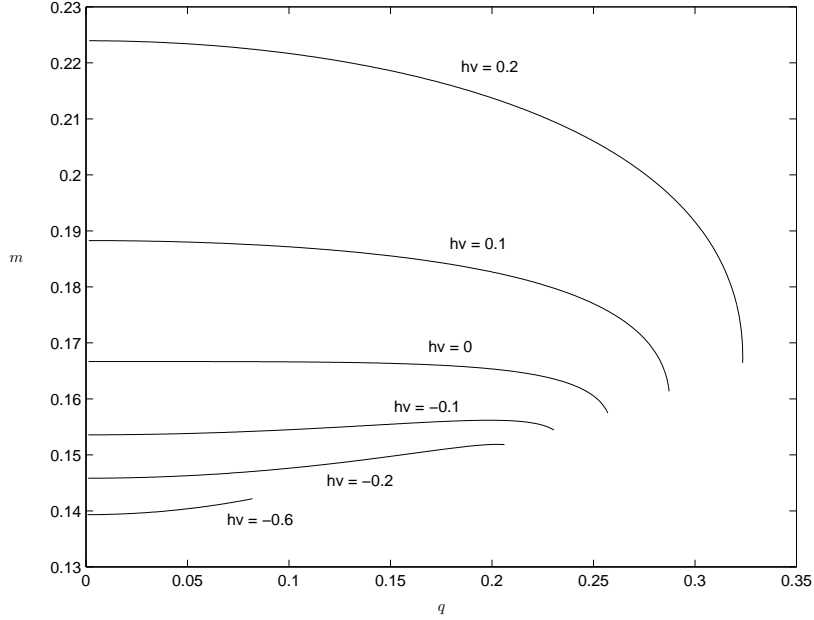}}
\put(-510,210){$m$}
\put(-240,-10){$q$}}
\end{center}
\caption{The dependence $m(q)$ for various $hv$. The mass $m$ is for
negative $q$ obtained by $m(q)=m(-q)$. The parameter $hKr_B$ satisfies
the energy conditions.\label{fig3}}
\end{figure}

The linear charge density of the shell as measured by an observer moving
with 4--velocity 
$\mathbf{u}_0=\partial _T/\sqrt{-g_{TT}}$ becomes equal to  
\begin{equation}\label{lin_charge_density}
q=2\pi\sqrt{g_{\Phi\Phi}}(-\mathbf{s}.\mathbf{u}_0)=
\frac{hKr_BX^{\frac{5}{2}}}{hv-1}\ .
\end{equation}
If we are given $v,r_B,h$ and want to find a dependence of some
quantity $R$ on $q$, we may express this dependence parametrically as 
$R=R(K), q=q(K)$. But a change in $K$ with $v,r_B,h$ kept constant
implies also a change of the mass parameters $\rho ,p$ 
and a change of the circumference of the shell $2\pi\sqrt{g_{\Phi\Phi}}$. 
Consequently one is not able to decide whether the changes of $R$ have
a physical origin in changes of the charge or the mass. 
However, the following equations hold 
\begin{equation}\label{changes}
\frac{\partial p}{\partial K}\mid _{K=0}=0\ ,\quad 
\frac{\partial\rho}{\partial K}\mid _{K=0}=0\ ,
\quad \frac{\partial}{\partial K}\sqrt{g_{\Phi\Phi}}\mid _{K=0}=0\
,\quad \frac{\partial q}{\partial K}\mid _{K=0}\neq 0\ .
\end{equation} 
Formulae \eref{changes} show that for small $K$ 
(and consequently for small $q$) the mass quantities expressed in terms
of functions of $q$ do not change significantly. 
The dependence $m(q)$, where 
$m=2\pi\sqrt{g_{\Phi\Phi}}\ t(\mathbf{u}_0,\mathbf{u}_0)$ is 
the linear mass density measured by an observer with the velocity
$\mathbf{u}_0$, is illustrated in figure \ref{fig3}.

\subsection{\label{CTC}Closed timelike curves}

In our spacetime the sign of the metric coefficient $g_{\phi\phi}(r)$
is crucial for existence or nonexistence of closed timelike curves
(CTC).  Our spacetime does not contain any horizons. This statement is
essential for further considerations and it will be proved in section
\ref{motion}. Because there are no horizons (generated by timelike
geodesics), if there exists a point $p$ with the property that
$g_{\phi\phi}(p)<0$, then for each point $q$ of the spacetime there
exists CTC which goes through $q$. However, each such curve must
intersect the region where $g_{\phi\phi}(r)<0$. 
If $g_{\phi\phi}(r)>0$ in the entire spacetime, there are no CTCs.
\begin{figure}[t] 
\begin{center}
\scalebox{0.6}{\includegraphics{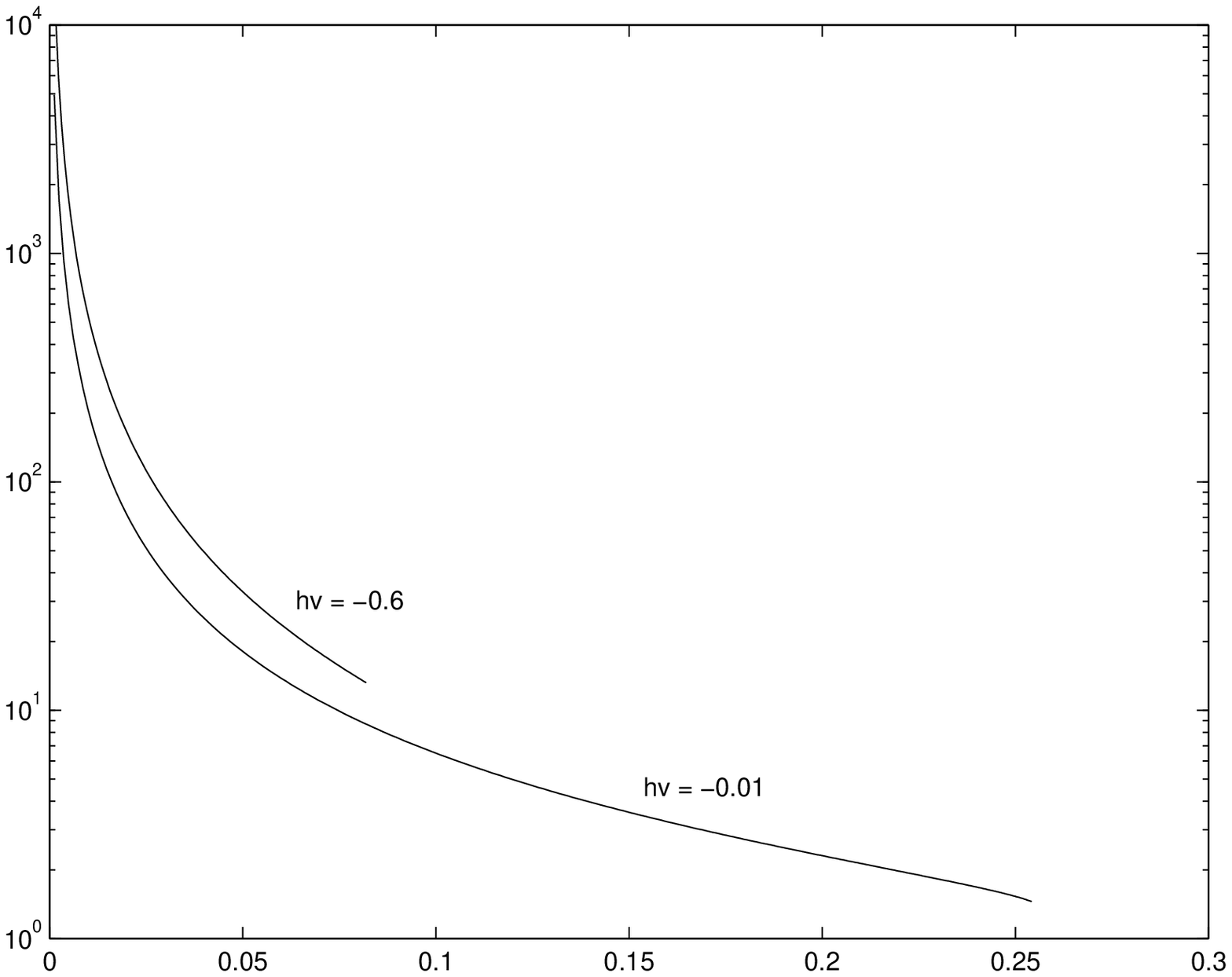}
\put(-510,210){$R_C$}
\put(-240,-10){$q$}}
\end{center}
\caption{Proper radial distance between $r_D$ and $r_C$ for $r_B=1$
and various $hv\leq0$. $R_C$ blows up when $q\rightarrow0$. It
corresponds to the fact that the 
uncharged limit contains no CTCs.\label{fig1}}
\end{figure}
 
It holds $g_{\phi\phi}(r)=(1-K^2r^2)r^2>0$ inside the shell, 
since it is always $r<\frac{1}{\mid K\mid }$ in BM spacetime. 
Outside the shell we have 
$g_{\phi\phi}(r)=-H^2(4r^2+\lambda r\ln (r))+2HDr$. To determine
the sign of $g_{\phi\phi}$, it is sufficient to examine the function 
$\frac{g_{\phi\phi}(r)}{r}$. One obtains  
\[\lim _{r\rightarrow\infty}\frac{g_{\phi\phi}}{r}=-\infty\ , 
\ \frac{g_{\phi\phi}}{r}(r_D)>0\ ,
\ \left(\frac{g_{\phi\phi}}{r}\right)^{\prime}
=-H^2(4+\frac{\lambda}{r})\ ,
\ \left(\frac{g_{\phi\phi}}{r}\right)^{\prime\prime}=
H^2\frac{\lambda}{r^2}\ ,\]
from which one can infer that there always exists 
$r_C\in(r_D,\infty)$ such that $g_{\phi \phi}>0$ for $r\in(r_D,r_C)$, 
$g_{\phi \phi}(r_C)=0$ and $g_{\phi \phi}<0$ for $r\in(r_C,\infty)$.  
Consequently, for each point $q$ of our spacetime and for all allowed
values of the free parameters there exists CTC which passes through
$q$. Nevertheless, all CTCs must intersect the region where $r > r_C$. 
If $hv>0$, then the same conclusion is true for uncharged limit of
spacetime, obtained by taking the limit $K\rightarrow 0$. On the other
hand, if $hv\leq 0$ and $K\rightarrow0$, 
then $g_{\phi\phi}(r)$ is positive in entire spacetime, so the
causality violation is avoided in this case.
\begin{figure}[h!]
\begin{center}
\scalebox{0.6}{\includegraphics{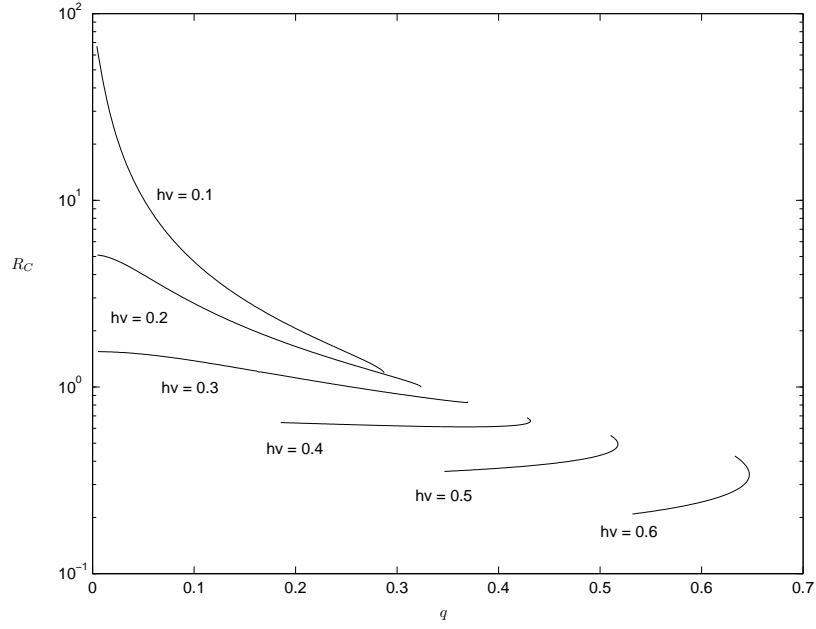}
\put(-510,210){$R_C$}
\put(-240,-10){$q$}}
\end{center}
\caption{Proper radial distance between $r_D$ and $r_C$ for $r_B=1$
and various $hv>0$. $R_C$ has a finite limit 
when $q\rightarrow0$. It corresponds to the fact that 
the uncharged limit contains CTCs too. \label{fig2}}
\end{figure}

The dependence of the proper radial distance $R_C=\int
^{r_C}_{r_D}\sqrt{g_{rr}}dr$ on the linear charge density $q$ is
depicted in figure \ref{fig1} in case $hv\leq 0, r_B=1$. The case $hv>0,
r_B=1$ is shown in figure \ref{fig2}. The quantities $R_C,q$ as functions of the free
parameters are of the form $R_C((hK)^2,r_B,hv)$ and $q(hKr_B,hv)$.
The curves in figure \ref{fig1} and figure \ref{fig2} plotted for given $r_B$ and
$hv$ are given parametrically. The parameter $hK$ runs through negative
values satisfying the conditions \eref{strong} and \eref{dominant}.
For given $hK$ the value of $q$ is computed from
\eref{lin_charge_density}, the value of $r_C$ is found by numeric
solution of the equation $g_{\phi\phi}(r)=0$. Since $R_C$ is even
function of $hK$ and $q$ is odd, positive values of $hK$ give curves
which differ from the ones in figure \ref{fig1} and figure \ref{fig2} only by the
substitution $[q,R_C]\rightarrow[-q,R_C]$.  Since
$R_C((hK/x)^2,xr_B,hv)=xR_C((hK)^2,r_B,hv)$ holds for $x>0$, the
curves for $r_B\neq 1$ can be obtained from the corresponding curves
in figure \ref{fig1} and figure \ref{fig2} by $[q,R_C]\rightarrow[q,r_BR_C]$.  

One can see that for small values of $| q|$, $R_C$ decreases with
increasing values of $|q|$, which is in a qualitative agreement with
\cite{Pavel}. 

\subsection{\label{electromagnetic}Electromagnetic field}

By normalizing vectors of the coordinate base inside the cylinder one
gets an orthonormal basis. The only non--zero independent component of
the interior electromagnetic field tensor $F$ in this base is
$F(\frac{\partial _r}{\sqrt{g_{rr}}},\frac{\partial _{\phi}}
{\sqrt{g_{\phi\phi}}})=B_{\hat{z}}=2K(1-K^2r^2)^2$.

Outside the cylinder we can construct the orthonormal basis 
\[\mathbf{v}=\frac{\sqrt{\mid\! g_{\phi\phi}\!\mid}(\partial _t
-\frac{g_{t\phi}}{g_{\phi\phi}}\partial _{\phi})}
{\sqrt{g_{t\phi}^2-g_{tt}g_{\phi\phi}}}\ ,\quad \mathbf{w}=\frac{\partial
_{\phi}}{\sqrt{\mid\! g_{\phi\phi}\!\mid}}\ ,\quad \mathbf{z}=
\frac{\partial _z}{\sqrt{g_{zz}}}\ ,\quad\mathbf{n}=\frac{\partial
_r}{\sqrt{g_{rr}}}\ .\]
Note that $g_{t\phi}^2-g_{tt}g_{\phi\phi}=(UD-H\Omega)^2r^2$ is always
positive. The vector field $\mathbf{v}$ is timelike for 
$g_{\phi\phi}>0$ while $\mathbf{w}$ is spacelike
\footnote{Observer with 4--velocity $\mathbf{v}$ is 
called ``locally nonrotating".}. 
If $g_{\phi\phi}<0$ the fields ${\bf v}$ and ${\bf w}$ will change their
roles. There are two non--zero independent components 
of the exterior electromagnetic tensor 
\[F(\mathbf{n},\mathbf{v})=hHr^{\frac{1}{4}}\frac{{\rm sign}
(g_{\phi\phi})}{\sqrt{\mid\! g_{\phi\phi}\!\mid}}\ ,\quad
F(\mathbf{n},\mathbf{w})=-\frac{Hr^{\frac{1}{4}}}{\sqrt{\mid\!
g_{\phi\phi}\!\mid}}\ .\] 
For $g_{\phi\phi}>0$ one has $E_{\hat{r}}=F(\mathbf{n},\mathbf{v})$ and $B_{\hat{z}}=F(\mathbf{n},\mathbf{w})$ and for $g_{\phi\phi}<0$ we have 
$E_{\hat{r}}=F(\mathbf{n},\mathbf{w})$ and
$B_{\hat{z}}=F(\mathbf{n},\mathbf{v})$. 
Both components vanish in infinity and their magnitudes diverge as $r$
approaches $r_C$. In the special relativity theory in order that the 
electromagnetic field be vanishing in infinity, one has to put the magnetic field 
outside the cylinder equal to zero, because of its homogeneity, but one 
can see that it is not the case in the general relativity. 

\section{\label{motion}Geodesic completeness and motion of test particles}
 
In this section we aim to study briefly a motion of test particles
that are generally charged, carrying a charge $e$. The trajectory is a
solution of the equation
\[\nabla _\mathbf{u}u^{\alpha}=-\nu F_{\beta}^{\;\;\alpha}u^{\beta}\ ,\] 
where constant $\nu$ stands for $\frac{e}m$ in case of a massive 
charged particle, while $\epsilon$ equals $1$ or $0$ depending on
whether we examine timelike or null curves (geodesics in the latter
case). The velocity vector field ${\bf u}=\frac{\d }{\d s}=
u^\alpha\partial_\alpha$ is assumed to be normalized, ${\bf
u}\cdot{\bf u}=-\epsilon$. We divide the discussion into two classes.

\subsection{Test particles in BM spacetime}

Because of the high degree of symmetry we have three conserved
quantities -- an energy $E$, an angular momentum $L$ and a momentum along 
$z$--axis $P_z$. These integration constants arise after first integration
of the equations of motion for a test particle of a mass $m$, carrying
a charge $e$. We find
\begin{subequations}
\begin{eqnarray}
\frac{\d t}{\d s}=E(1-K^2r^2)^2\ ,\label{MBt}\\
\frac{\d\phi}{\d s}=\frac{L-K\nu r^2}{r^2(1-K^2r^2)}\ ,\label{MBfi}\\
\frac{\d z}{\d s}=P_z(1-K^2r^2)^2\ .\label{MBz}
\end{eqnarray}
\end{subequations}
Radial coordinate fulfills the following equation
\begin{equation}
\frac{1}2\left(\frac{\d r}{\d s}\right)^2=V_{\rm eff}\ ,\label{MBeff} 
\end{equation}
with the effective potential $V_{\rm eff}$ given by
\begin{eqnarray}\label{effective}
V_{\rm eff}=\frac{1}{2r^2}(1-K^2r^2)^4
\left[\left(E^2-P_z^2\right)r^2\left(1-K^2r^2\right)^3-\right.\nonumber\\
\left.-\epsilon r^2\left(1-K^2r^2\right)-\left(L-K\nu r^2\right)^2\right].
\end{eqnarray}

The motion of a test particle is clearly restricted to values of $r$
for which \eref{effective} (and consequently the expression in square
brackets of \eref{effective}) is non--negative. First of all this
means that $F^2\equiv E^2-P_z^2\geq0$.
By using the estimate $x(1-x)\leq\frac{1}4$ for $0\leq x\leq 1$,
where $x\equiv K^2r^2$, one can make a rough conclusion that a 
motion of a test particle is forbidden at least for the following
ranges\footnote{This is an estimate, the exact ranges are wider.} 
of the integration constants ($G\equiv KL$):
\begin{itemize}
\item [i] for $F^2\leq4(G-\nu)^2+\epsilon$\quad if \quad $G\geq\nu\geq0$ 
\quad or \quad \n 0{\nu}G.
\item [ii] for $F^2\leq4G^2+\epsilon$\quad if \quad \n G0{\nu} 
\quad or \quad \n {\nu}0G.
\item [iii] for $F^2\leq\epsilon$\quad if \quad \n {\nu}G0
\quad or \quad \n 0G{\nu}.
\end{itemize}

\begin{figure}[t]
\begin{center}
\scalebox{0.7}{\includegraphics*[32mm,96mm][20cm,20cm]{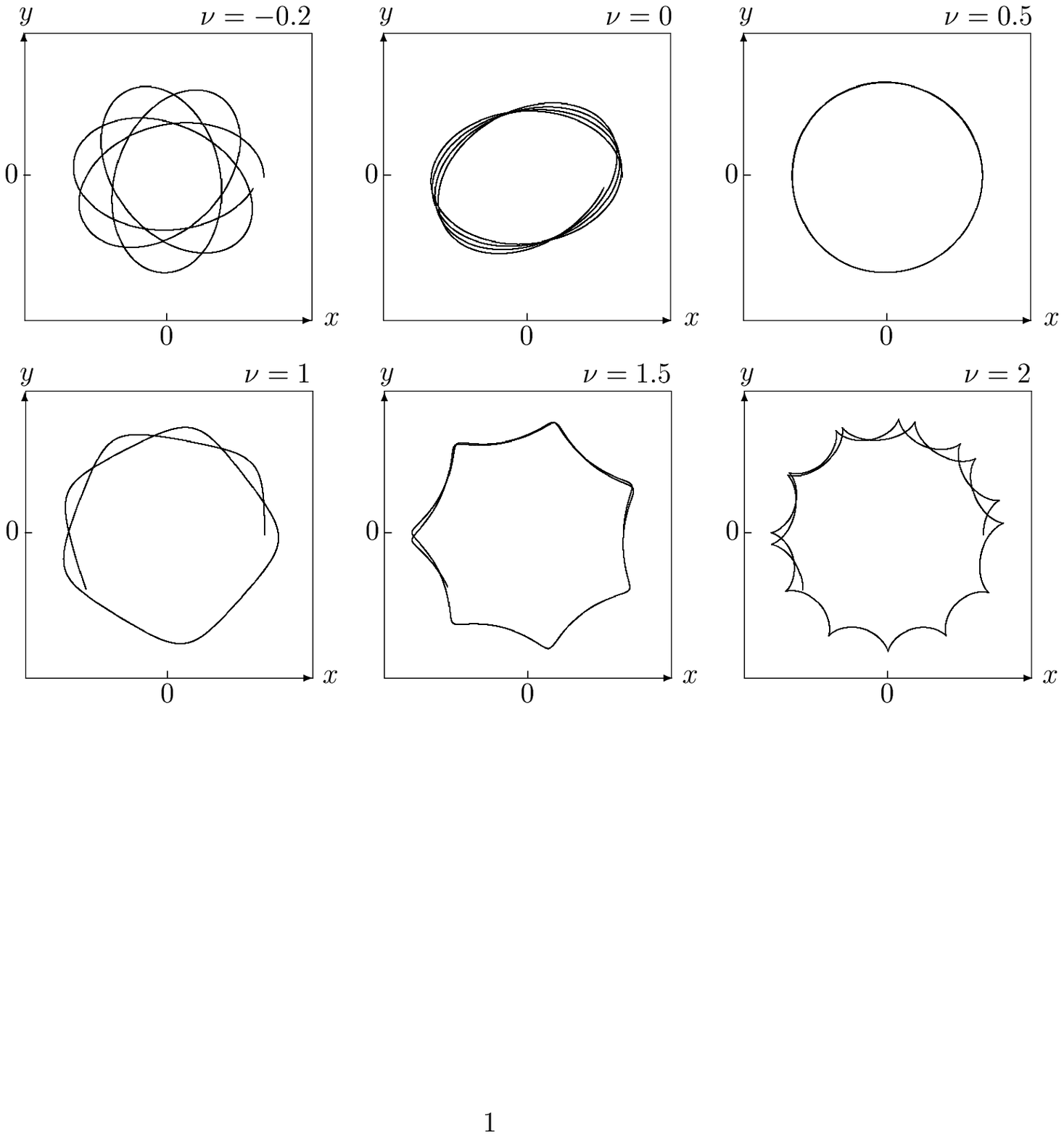}}
\end{center}
\caption{\label{fig4}
Some trajectories of charged test particles inside the cylinder with
$z={\rm const}$. The coordinates $x,y$ depend on $r,\phi$ as usually: 
$x=r\cos\phi,\ y=r\sin\phi$. The initial values and the value of $K$ 
are same for all the pictures. In the natural units we have 
$\frac{\d r}{d\tau}(0)=\frac{\d z}{d\tau}(0)=0,\
\frac{\d\phi}{d\tau}(0)=1,\ r_0=0.1,\ K=1$. The values of $\nu$ are
written in the figure.}
\end{figure}

Although it is not generally possible to integrate \eref{effective}
in terms of elementary functions, one can obtain an exact expression for the angular
velocity $\omega\equiv \frac{\d \phi}{\d \tau}$ of a test particle
moving on a circular trajectory $z={\rm const}$, $r=r_0$.
In units SI it reads
\begin{equation}\label{frek}
\omega =\frac{K}{1-4\mathcal{G}K^2r_0^2}
\left(-\nu\pm\sqrt{\nu ^2+2\mathcal{G}c^2\frac{1-4\mathcal{G}K^2r_0^2}
{1-\mathcal{G}K^2r_0^2}}\right)\ ,
\end{equation}
where $\mathcal{G}=\frac{4\pi G}{\mu _0c^4}\approx 8,15.10^{-38}\, 
T^{-2}m^{-2}$. 
If the gravity is switched off ($G=0$) one obtains the classical 
cyclotron frequency $\omega _{STR}=-2K\nu =-\frac{B_ze}{m}$, since it is
$B_z=2K$ for $G=0$. Expanding \eref{frek} in $\mathcal{G}$ we 
get the approximate expression for the relative deviation of cyclotron 
frequency caused by gravity 
\[\frac{\omega  -\omega _{STR}}{\omega _{STR}}\approx
\mathcal{G}(\frac{c^2}{2\nu ^2}+4K^2r_0^2)\ .\]
To see that the gravity has only a small influence on cyclotron
frequency in usual conditions, let us evaluate how much $\omega$ differs from
$\omega_{STR}$. For an electron, $K=10 T$ and $r_0=0,1m$, one has
the relative deviation of order $10^{-37}$. Some trajectories of the charged 
test particles are depicted in figure \ref{fig4}.

\subsection{\label{DRmotion}Test particles in DR spacetime}
For the sake of technical simplicity, we will consider the line
element \eref{MC_puv}.
Just like in BM spacetime we are led to the following equations of
motion (with bars over coordinates omitted)
\begin{subequations}
\begin{eqnarray}
\frac{\d t}{\d s}&=&\frac{L}r\ ,\label{t}\\
\frac{\d \phi}{\d s}&=&\nu -\frac{E}r+L(4+\lambda\frac{\ln r}r)\
,\label{fi}\\
\frac{\d z}{\d s}&=&\sqrt rP_z\ ,\label{z}
\end{eqnarray}
\end{subequations}
along with the following effective potential governing the 
radial motion of a test particle
\begin{equation}
 V_{\rm eff}=-\frac{1}2\sqrt r\left[4L^2+\epsilon+2\nu L+\sqrt
rP_z^2-\frac{2E}rL+L^2\lambda \frac{\ln r}r\right]\ .
\label{MCeff}
\end{equation}
Since an integration of \eref{MCeff} is not possible explicitly in
general, we will discuss here a basic behavior only. First of all,
irrespective of values of the integration constants, $V_{\rm eff}$
tends to minus infinity as $r$ goes to infinity. This means that the test
particles radial motion is bounded and it is forbidden behind a
certain critical radius depending on a particular path.

On the other hand, if $\lambda>0$ there is $r_1$ such that $V_{\rm
eff}>0$ for $r<r_1$. If $\lambda<0$ the situation gets more
complicated depending on values of others integration constants. 
In this case both possibilities can occur. Either a range $r\in[r_0,r_1]$
exists in which a test particle motion is permitted, or it may become
that the motion will be forbidden entirely.
We shall not be concerned with an investigation under which circumstances 
either behavior
could occur further. Instead, provided the test particles can move and 
the inequality $r_1>r_D$ is satisfied, we will show that the spacetime
is complete. 

Because according to our assumption $r$ is bounded both from bellow
and above, it follows from the equations \eref{t}-\eref{z} that the terms
$\frac{\d t}{\d s},\ \frac{\d \phi}{\d s}$ and $\frac{\d z}{\d s}$ can
be bounded by suitable chosen constants $A,B$ and $C$ as $\left|\frac{\d z}{\d
s}\right|<A$ and so on for $t$ and $\phi$. Thus the test particle can
escape to infinity in $z$ or $t$ direction only in infinite $s$
showing that every geodesic\footnote{The geodesics are obtained by 
setting $e$ equal to
zero.} in our matched spacetime can be continued to an arbitrary value. 

Note that the component $g_{00}$ of the metric tensor \eref{MC_puv}
vanishes for $r=0$. Moreover, if $\lambda$ is positive one has one
additional root, and if $\lambda<-4{\rm e}$ (${\rm e}$ is Euler
constant), we have two additional 
roots of $g_{00}$. As we shall see immediately, $r=0$ represents the
physical singularity while the two additional roots do not. 
Since in stationary spacetimes frequencies of a light signal 
measured at two distinct points $p$ and $q$ of the geodesic along 
which the signal moves are related by
$\frac{\omega(p)}{\omega(q)}=\sqrt{\frac{g_{00}(q)}{g_{00}(p)}}$, one
finds the physical interpretation of these two roots: they corresponds
to hypersurfaces of infinite red shift. Being timelike, these
hypersurfaces are not horizons.

A direct computation of the Riemann curvature tensor invariants shows
that the rotation axis $r=0$ is the only intrinsic singularity. For
instance the first non--trivial\footnote{The simplest invariant, Ricci
curvature, clearly vanishes identically and it also holds for
$R_{\mu\nu}R^{\mu\nu}$.} curvature invariant is equal to
\[R_{\alpha \beta \gamma \delta}R^{\alpha \beta \gamma \delta}=
\frac{3}{4r^3}\ .\]
Generally, dangerous terms involved in the components of the Riemann tensor, 
that could be possibly responsible for singularities, are proportional to
$r^{k_1}$, $r^{k_2}\ln r$, where the constants $k_1$ and $k_2$ acquire
negative values in all but one case, namely when $k_1=1/2$. It can be seen
that their combination resulting from a computation of any Riemann
curvature invariant is tame for $r$ going to infinity. Thus the
rotation axis constitutes the only physical singularity, but because
we have truncated DR spacetime on $r_D$ and restricted ourselves to
region $r\in[r_D,\infty)$, we are left with entirely non--singular
spacetime.

\section{\label{conclusion}Conclusion}

We found spacetime, where infinite rotating cylindrical shell from
charged perfect fluid acts as a source.  Bonnor--Melvin magnetic
universe has been used as the interior part of the cylinder and Datta
and Raychaudhuri spacetime as the exterior one.  Because of the
junction conditions, the metric, the electromagnetic potential and the
shell parameters have been expressed as functions of four free
parameters $K,r_B,v,h$, where $K$ is closely connected with the
magnetic field inside the shell, $r_B$ is the value of the radial
coordinate at witch the interior spacetime has been cut off (i.e.
where the shell is located), $v$ is $\Phi$--component of the shell
particles velocity and $h$ can have only values $\pm 1$. The question
was examined in what ranges of the parameters the energy conditions
are satisfied. 

The spacetime found contains closed timelike curves for all allowed
values of the free parameters. Also it was shown that the radial
distance between the shell and the radius $r_C$ behind which each CTC
must pass, decreases with an increasing absolute value of the linear
charge density $| q|$ at the shell for small values of $| q|$.
Finally an investigation of the test particles trajectories, 
either charged or uncharged,
was carried out both for the interior as well as exterior region. 
In particular it was found the the trajectories are always radially
bounded and that the resulting spacetime is free of physical
singularities.

\section*{Acknowledgments}
This work was supported by grant 201/03/0512. One of us (PK) also
wishes to acknowledge to grant 202/03/P113 of Grant Agency of the Czech
Republic.


\end{document}